\documentstyle[twoside,fleqn,espcrc2,epsf]{article}

\newcommand{\AmS}{{\protect\the\textfont2
  A\kern-.1667em\lower.5ex\hbox{M}\kern-.125emS}}
\newcommand{\To}{\rightarrow}
\newcommand{\life}{\tau_{\tau}}
\newcommand{\ee}{e^+e^-}
\newcommand{\mumu}{\mu^+\mu^-}
\newcommand{\tautau}{\tau^+\tau^-}

\newcommand{\gaga}{\gamma\gamma}
\newcommand{\qqbar}{\textsl{q}\bar{\textsl{q}}}
\newcommand{\mrad}{\,{\mathrm{mrad}}}

\newcommand{\mic}{\,\mu{\mathrm{m}}}

\newcommand{\cm}{\,{\mathrm{cm}}}

\newcommand{\fs}{\,{\mathrm{fs}}}
\newcommand{\gev}{\,{\mathrm{GeV}}}

\newcommand{\oneone}{\mbox{1-1}}
\newcommand{\onethree}{\mbox{1-3}}
\newcommand{\threethree}{\mbox{3-3}}
\newcommand{\mylists}{\setlength{\leftmargini}{7mm}%
                      \setlength{\labelsep}{1.7mm}}
\title{Review of $\tau$ lifetime measurements}

{\catcode`@=11\gdef\@makeadmark#1{\hbox{}}}  

\author{Steven R.\ Wasserbaech%
\address{Department of Physics, P.O.\ Box 351560,
University of Washington, Seattle WA 98195-1560 USA}
\thanks{Supported by U.S.\ National Science Foundation grant PHY-9605225.}}
       
\begin{document}
\begin{abstract}
The measurements of the mean lifetime of the $\tau$ lepton are reviewed.
The conditions for measuring the lifetime
at various $\ee$ colliders are compared
and the analysis methods are briefly described.
The new developments since the previous 
Workshop on Tau Lepton Physics are listed.
The world average is $\life = 290.5 \pm 1.0 \fs$.
The LEP experiments dominate this average and have
analyzed nearly all of their data.
In anticipation of the next era of precision measurements
at CLEO and the b factories,
the important sources of systematic errors
and the treatment of systematic biases are discussed.\\[7mm]
\textit{Invited talk presented at the
Fifth International Workshop on Tau Lepton Physics,\\
Santander, Spain, 14--17 September 1998}\\[2mm]
\end{abstract}

\maketitle

\section{Introduction}

The mean lifetime of the $\tau$ lepton is deduced from
geometrical reconstruction of $\tau$ daughter tracks
in $\ee\To\tautau$ events.
We need to know the $\tau$ lifetime in order to make
certain tests of lepton universality.
Such tests are sensitive to the $\tau\nu_{\tau}{\rm W}$ coupling
and also to possible new physics~\cite{univ};
at present the precision is limited by
the experimental uncertainties on
the $\tau$ lifetime and branching fractions.
The $\tau$ lifetime is also useful for
evaluating the strong coupling constant 
$\alpha_{\rm S}$~\cite{alphas}.

In this talk I use the conventional coordinate system
for $\ee$ experiments,
i.e., with the $z$ (polar) axis along the direction of
the incident beams.
The impact parameter $d$ of a reconstructed charged track 
is measured in the $xy$ projection
with respect to the nominal interaction point;
$d$ is signed according to the $z$ component of the track's
angular momentum about this point.
The azimuthal decay angle $\psi$ of a $\tau$ daughter track is defined
as the signed quantity $\phi_{\rm daughter} - \phi_{\tau}$
in the laboratory frame,
where $\psi$ lies between ${-}\pi$ and $\pi$.

\begin{figure}[t]
\begin{center}
\mbox{\epsfysize=94mm\epsffile{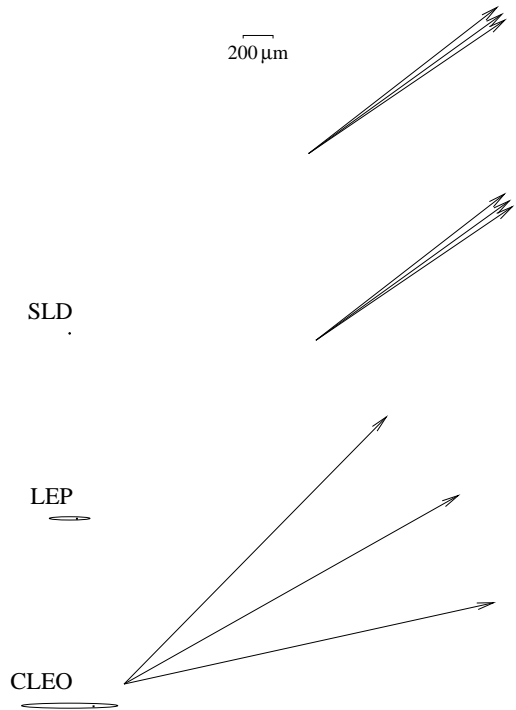}}%
\end{center}\par\vspace{-10mm}
\caption{Experimental conditions at SLD, LEP, and CLEO.
All drawings show the same three-prong $\tau$ decay,
boosted to the appropriate energy for each experiment and
projected onto the $xy$ plane ($20{\times}$ actual size).
The ellipses representing the luminous region include the
typical uncertainty on the beam axis coordinates.}
\label{f:cond}
\end{figure}

At present, useful measurements of the $\tau$ lifetime can be obtained
by SLD, the four LEP experiments, and CLEO,
and the experimental conditions are quite different in these places.
As far as the lifetime measurement is concerned, the relevant 
differences include the $\tau$ flight distance in the laboratory
(which varies as $\beta\gamma$ of the $\tau$ boost),
the opening angles of the $\tau$ decay products 
(which vary roughly as $1/\gamma$),
the dimensions of the luminous region,
and the number of collected $\tau$ pairs.
On the other hand, 
the impact parameters of the $\tau$ daughters are on the order
of $c\life = 87\mic$ in all experiments,
as long as the $\tau$ has $\beta \sim 1$.
Comparisons of the experimental conditions are given in 
Table~\ref{t:cond} and Fig.~\ref{f:cond}.

\section{Methods}
In this section I give a brief description of the $\tau$ lifetime
measurement methods that are currently in use.

\subsection{Decay length method}
The decay length (DL) method (or vertex method)
is applied to three-prong $\tau$ decays~\cite{markii}.
The procedure consists of reconstructing the $\tau$ decay vertex
in two or three dimensions.
In order to evaluate the $\tau$ decay length, an estimate of the
$\tau$ production point is also needed.
In an $\ee\To\tautau$ event, the only available estimate is
the centroid of the luminous region.
Because the luminous region is huge
along the direction of the beam axis
(considerably larger than the typical $\tau$ flight distance),
we can effectively estimate only two of the three coordinates
of the $\tau$ production point,
and we end up with a measurement of the $\tau$ displacement $L_{xy}$
in the $xy$ projection.
An estimate of the polar angle $\theta_{\tau}$ of the $\tau$ direction,
taken for example from the event thrust axis,
is therefore needed in order to calculate the
$\tau$ flight distance $L$ in three dimensions.
The mean lifetime is deduced from the mean of the
$L$ distribution
and the mean value of $\beta\gamma$;
the latter is calculated from simulated $\tautau$ events,
including initial and final state radiation,
after the event selection criteria are applied.

\begin{figure}[t]
\begin{center}
\mbox{\epsfxsize=42mm\epsffile{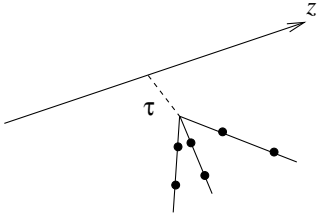}}%
\end{center}\par\vspace{-10mm}
\caption{A $\tau$ decay in which the three charged daughter tracks
emerge in the same azimuthal direction $\phi$.
The vertex cannot be reconstructed from the track measurements
in the $r\phi$ view alone,
but a full three-dimensional vertex reconstruction is possible when
the measurements in the $rz$ view are added.}
\label{f:rzview}
\end{figure}

\begin{table}[t]
\caption[]{Experimental conditions for $\tau$ lifetime measurements.
Here, $N_{\tau\tau}$ is the approximate number of produced 
$\ee\To\tautau$ events in the data sample,
$E_{\tau}$ is the $\tau$ energy in the laboratory
(neglecting radiative effects),
$\gamma$ refers to the boost of the $\tau$,
and $\beta\gamma c\life$ is the mean $\tau$ flight distance
in the lab.}
\label{t:cond}
\begin{tabular}{@{}lrrr}\hline
                              & \makebox[13mm]{\hfill SLD}  
                              & \makebox[13mm]{\hfill LEP}  
                              & \makebox[13mm]{\hfill CLEO} \\ \hline
$N_{\tau\tau}$                & 20K  & $4\times200$K & 5M\rlap{${}^{\rm a}$} \\
$E_{\tau}$ (GeV)              & 45.6 & 45.6          & 5.3                   \\
$\gamma$                      & 25.7 & 25.7          & 3.0                   \\
$\beta\gamma c\life$ ($\mu$m) & 2200 & 2200          & 240                   \\
\hline
\multicolumn{4}{@{}l}{\small ${}^{\rm a}$%
\parbox[t]{68mm}{\small Collected with the present
(CLEO 2.5) detector configuration.}} \\
\end{tabular}
\end{table}
Modern vertex detectors can measure precise impact parameters in
both the $r\phi$ and $rz$ views.
Because the luminous region is so large along $z$,
the possibility of measuring the $z$ coordinate of the $\tau$ decay
vertex is not particularly useful for the classical DL method.
However, it is important to realize that the tracking information
from the $rz$ view can also significantly improve the measurement of 
the $(x,y)$ coordinates of the decay vertex,
from which the lifetime is extracted.
This point is illustrated in Fig.~\ref{f:rzview}.

At LEP and SLD, the size of the luminous region and the tracking
resolution are such that the statistical uncertainty on the mean
decay length is dominated by the natural width of the exponential $t$
distribution;
the relative uncertainty on the mean lifetime is not far from
its optimum value, $\Delta\life/\life = 100\% / \sqrt{N_{\tau}}$,
where $N_{\tau}$ is the number of selected $\tau$ decays.
The systematic errors in DL analyses at high energies
also tend to be fairly small.
In short, our detectors and our technique
for measuring the $\tau$ lifetime 
from three-prong decays at LEP and SLD
are very effective.

At lower center-of-mass energies the $\tau$ decay length is shorter
and the size of the luminous region can significantly dilute the
precision of the measurement.
In such cases the decay length resolution can be improved by
considering the separation of the two decay vertices in
\threethree\ topology events.
This procedure eliminates the smearing due to the size 
of the luminous region, 
but it does not dramatically improve the final results because
the statistics are low and the $\qqbar$ background is larger
in the \threethree\ channel.

\subsection{One-prong decays}
While the DL method is entirely satisfactory for analyzing three-prong
$\tau$ decays,
we cannot apply such a straightforward technique to the one-prong decays.
Due to the unobserved neutrinos, the $\tau$ direction is unknown.
We cannot reconstruct the decay length of an individual $\tau$ decaying
into one prong.
We can nevertheless measure the mean decay length from a collection
of one-prong decays.
The relative statistical uncertainty will be considerably larger than
$100\%/\sqrt{N_{\tau}}$.
We will do best if we analyze both $\tau$ decays in an event together.

It is difficult to incorporate all of the event information into
a simple method.
We now use several methods to analyze one-prong decays.
Each method uses a different subset of the available information, 
and none of the methods is vastly superior to the others.
We combine the results from the various methods,
taking into account the statistical and systematic correlations,
to utilize as much of the available information as possible.
I now proceed to describe these methods in more detail.

\subsection{Impact parameter method}
The impact parameter (IP) method
is applied to one-prong $\tau$ decays~\cite{mac}.
In this method, an estimate of the $\tau$ direction is taken,
for example, from the event thrust axis.
The lifetime-signed impact parameter of the daughter track is
then defined:
\[ D = \left\{ \begin{array}{ll}
d   & \ \ \mbox{if $\psi > 0$;} \\
-d  & \ \ \mbox{if $\psi < 0$.}
\end{array} \right. \]
The mean of the $D$ distribution is then roughly proportional
to $\life$.
The dependence of the $D$ distribution on $\life$ is determined
from Monte Carlo simulation.

Decays with $D<0$ result from
the impact parameter resolution,
the size of the luminous region,
and errors on the $\tau$ direction.
The last of these effects is probably the most dangerous because
it brings about a substantial change in the mean of $D$,
and we rely on the Monte Carlo to correctly describe the
$\tau$ direction errors.
Although we tend to think of the IP method as a one-$\tau$-at-a-time
method, 
the $\tau$ in the opposite hemisphere contributes to the
thrust axis determination 
and hence affects the $D$ distribution.

\subsection{Impact parameter difference method}
\label{ss:ipd}
The impact parameter difference (IPD) method
is applied to \oneone\ topology events~\cite{alephipd}.
In this method the mean $\tau$ decay length is extracted
by considering the correlation between the difference on the
daughter track impact parameters $d$ and the
difference of their azimuthal angles $\phi$.
Specifically, we define
$Y = d_+ - d_-$ and
$X = \Delta\phi \sin\theta_{\tau}$,
where $\Delta\phi = \phi_{+} - \phi_{-} \pm \pi$
is the acoplanarity of the two daughter tracks.
If the $\tau^{+}$ and $\tau^{-}$ are back to back
in the $xy$ projection 
and the decay angles $\psi$ are small, 
such that $\sin\psi \cong \psi$,
we find, at a particular value of $X$, that
$\langle Y \rangle = \bar{L}X$,
where $\bar{L}$ is the mean $\tau$ decay length in the lab.
A fit to the $Y$~vs $X$ distribution is performed to
extract the slope $\bar{L}$.
The polar angle $\theta_{\tau}$ is taken from the
event thrust axis;
the resulting error on the $\tau$ direction has a negligible
effect on the fitted $\bar{L}$.
The main disadvantage of the IPD method is that 
the uncertainty on the $\tautau$ production point due to the
size of the luminous region enters twice in the smearing on $Y$.

\subsection{Impact parameter sum methods}
The miss distance (MD)~\cite{delphimd} and
momentum-dependent impact parameter sum (MIPS)~\cite{alephmips}
methods
are designed to give improved statistical precision
by virtually eliminating the smearing effects related
to the size of the luminous region.
In a \oneone\ topology event we define the ``miss distance'' 
$\Delta = d_{+} + d_{-}$.
This sum of impact parameters is, roughly speaking,
the distance in the $xy$ projection between the two daughter tracks
at their closest approach to the beam axis.
This quantity is almost independent of the $\tautau$ production point.
The $\Delta$ distribution depends on $\life$;
a Monte Carlo simulation is used to parametrize the true distribution,
in order to extract the lifetime from the data.
The main disadvantage of these methods is that 
the results of the fit to the data are sensitive to the
assumed impact parameter resolution.

I refer to the simplest form of this analysis as the MD method.
The MIPS method is a refinement of MD in which the $\Delta$ distribution
is parametrized in terms of the momenta in the lab of the
two $\tau$ daughter tracks.
DELPHI's new results announced at this workshop feature the MIPS
refinement and one other: the $\Delta$ distribution is parametrized
separately for lepton and hadron daughters;
I refer to this method as MD++.

\begin{figure}[b]
\begin{center}
\mbox{\epsfxsize=30mm\epsffile{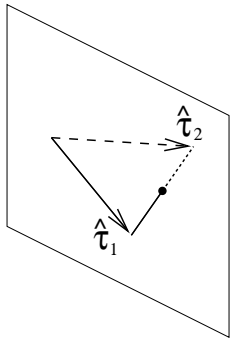}}
\end{center}\par\vspace{-10mm}
\caption{In the 3DIP method,
the event is projected along the direction
given by $\hat{\tau}_2 - \hat{\tau}_1$,
where $\hat{\tau}_{1,2}$ are the
two possible $\tau^{-}$ momentum directions.}
\label{f:threedip}
\end{figure}
\subsection{Three-dimensional impact parameter method}
The three-dimensional impact parameter\linebreak 
(3DIP) method
makes use of more of the kinematic information in the events 
to yield a higher sensitivity per event than the other
one-prong methods~\cite{alephthreedip}.
The main disadvantage is that the method can only be applied to
{\sl hadron\/} vs {\sl hadron\/} events
(42\% of all $\tautau$ events).
Because a $\tau\To{\sl hadron}$ decay yields only one
unobserved neutrino,
it is possible to reconstruct the $\tau$ direction
in {\sl hadron\/} vs {\sl hadron\/} events
up to a twofold ambiguity.
Let $\hat{\tau}_1$ and $\hat{\tau}_2$ denote the two
possible $\tau^{-}$ directions
reconstructed for a particular event (Fig.~\ref{f:threedip}).
If we then project the event along a direction
chosen such that $\hat{\tau}_1$ and $\hat{\tau}_2$
coincide,
we end up with no uncertainty on the $\tau$ direction
in that projection.
We then define a generalized impact parameter sum in that projection,
so that there is almost no smearing due to the size of the luminous
region.
A fitting procedure operates on this impact parameter sum and
on the two projected $\tau$ decays angles in order to
extract the mean lifetime.

The 3DIP method is the first to use impact parameter information
from the $rz$ view in the analysis of one-prong decays.
The method has the extremely important advantage that the
tracking resolution and the $\tau$ lifetime can be extracted
simultaneously from the $\tautau$ events.

\section{New lifetime results since TAU96}
There are four new developments to report:
{\mylists
\begin{itemize}
\item
At TAU98, L3 is reporting preliminary results from their
1994--95 data, analyzed with the DL and IP methods~\cite{ltrois}.
The new L3 average (1991--95 data) is 
$\life =
(291.7 \pm 2.0 \,[{\rm stat}] \pm 1.8 \,[{\rm syst}])\fs$.

\item
At TAU98, DELPHI is reporting preliminary results from their
1994--95 data, analyzed with the DL, IPD, and
MD++ methods~\cite{delphi}.
The new DELPHI average (1991--95 data) is
$\life =
(291.9 \pm 1.6 \,[{\rm stat}] \pm 1.1 \,[{\rm syst}])\fs$.

\item
The thesis of Patrick Saull (ARGUS)~\cite{saull} describes the
vertex impact parameter (VIP) method,
which provides improved lifetime sensitivity
for \onethree\ topology events
in cases where the size of the luminous region limits the
precision of the DL method.
The VIP method uses the impact parameter of the one-prong
track with respect to the three-prong vertex,
and the acoplanarity of the one- and three-prong jets.
(A similar approach is described in~\cite{opalvip}.)

\item
In 1997, ALEPH published results from 
the 3DIP method (1992--94 data)~\cite{alephthreedip}
and from the MIPS, IPD, and DL methods (1994 data)~\cite{aleph},
preliminary versions of which had been shown at TAU96.
\end{itemize}}

\section{Summary of measurements}
\begin{figure}[b]
\begin{center}
\mbox{\epsfxsize=60mm\epsffile{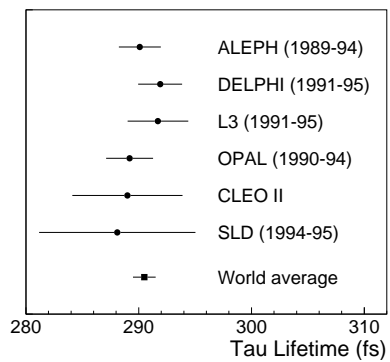}}%
\end{center}\par\vspace{-10mm}
\caption{Measurements of the $\tau$ lifetime.}
\label{f:summary}
\end{figure}

In calculating the world average $\tau$ lifetime,
I follow the Particle Data Group~\cite{pdg} and
ignore early measurements with large uncertainties.
The measurements are listed in Table~\ref{t:summary}
and plotted in Fig.~\ref{f:summary}.
In most cases the results shown are themselves averages
of two or more measurements obtained by a given
experiment with different methods and/or data samples.
\begin{table}[t]
\setlength{\tabcolsep}{1.5pc}
\caption[]{Measurements of the $\tau$ lifetime.}
\label{t:summary}
\begin{tabular}{@{}lr@{$\,\pm\,$}c@{$\,\pm\,$}l}
\hline
Experiment & 
\multicolumn{3}{c}{$\life\pm\mbox{stat}\pm\mbox{syst}$ (fs)} \\ \hline
ALEPH~\cite{alephthreedip,aleph}  & 290.1 & 1.5 & 1.1 \\
DELPHI~\cite{delphi}${}^{*}$      & 291.9 & 1.6 & 1.1 \\
L3~\cite{ltrois}${}^{*}$          & 291.7 & 2.0 & 1.8 \\
OPAL~\cite{opal}                  & 289.2 & 1.7 & 1.2 \\
CLEO~II~\cite{cleo}               & 289.0 & 2.8 & 4.0 \\
SLD~\cite{sld}${}^{*}$            & 288.1 & 6.1 & 3.3 \\ \hline
\multicolumn{4}{@{}l}{${}^{*}$Preliminary} \\
\end{tabular}
\end{table}

The world average is
$\life = 290.5 \pm 1.0 \fs$,
where the systematic errors in the various experiments are
assumed to be uncorrelated.
The $\chi^2$ describing the consistency of the measurements is
1.36 for 5 degrees of freedom,
corresponding to a confidence level of 0.929.
The four LEP experiments contribute 94\% of the total weight
in the average.
Since the beginning of the LEP era, the uncertainty on the world
average has been reduced by a factor of 8.
By now, almost all of the LEP1 data has been analyzed.
(The LEP2 data is not expected to yield 
any useful $\tau$ lifetime results.)

\section{Systematic errors}
Most recent measurements of the $\tau$ lifetime are statistics limited.
Nevertheless, it is useful to examine the systematic effects that 
we will need to deal with in the coming years in order to further
improve the precision of the measurements.
Some of the important sources of systematic errors are
tracking errors (simulation and/or parametrization);
vertex fit and lifetime extraction procedures;
and detector alignment.
I will now discuss each of these topics in turn.

\subsection{Tracking errors}
I would like to mention two delicate issues related to 
tracking errors.

The first issue concerns the dependence, in some methods,
of the measured lifetime on the assumed impact parameter resolution.
In such cases, it is mandatory to measure the resolution
from reconstructed tracks in the real data.
Bhabha and dimuon events are readily available for this job,
but the high momentum electrons and muons in those samples
are not representative of the $\tau$ daughter tracks and
the contribution to the impact parameter resolution
from multiple scattering cannot be studied.
Some experiments employ $\gaga\To\ee$, $\mumu$ events
to parametrize the resolution at the low end of
the momentum range.
While these test samples can give a fairly precise description
of the impact parameter resolution for electrons and muons,
it is not easy to use the real data 
to parametrize the effects of nuclear interactions
on pion and kaon tracks.

The second issue concerns the correlation between the errors
on the reconstructed impact parameter and direction of a track,
e.g., between $d$ and $\phi$.
This correlation is positive and results from the extrapolation
of the reconstructed tracks 
from the measured points in the tracking detectors 
to the interaction region.
In some methods the correlation can simulate a longer $\tau$ 
lifetime~\cite{srw}.
The effect is especially bad for the $\tau$ (compared to other
particles) because 
(1) the short $\tau$ lifetime leads to small impact parameters,
(2) the small $\tau$ mass leads to small decay opening angles
(which get smaller at higher $\sqrt{s}$), and
(3) we fit to the entire proper time spectrum
(as opposed to the situation in charm lifetime measurements
in fixed target experiments,
where a cut $L > L_{\rm min}$ is imposed and the
mean lifetime is determined from the {\it slope\/} of the proper time
distribution).
The effects of the tracking errors on the measured $\life$
must be carefully taken into account.

\subsection{Vertex fit and lifetime extraction\\ procedures}
Although the direct reconstruction of $\tau$ decay lengths
in the DL method appears to be quite straightforward,
several subtle effects are present,
yielding biases on the measured lifetime.
These effects are related to the tracking resolution,
and they can be substantial in experiments where the
vertex resolution along the $\tau$ direction is larger
than the mean decay length
(not the case at SLD and LEP).
{\mylists
\begin{itemize} 
\item Due to the correlation between 
the track impact parameter and direction errors,
fluctuations to larger opening angles in a three-prong decay
tend to be associated with upward fluctuations
in the reconstructed decay length.
The larger opening angles also lead to 
a smaller calculated uncertainty on $L$.
Thus the upward fluctuations in $L$ are associated with larger weights
in the calculation of the mean decay length.
\item Radiative events have lower $\tau$ momenta,
which tend to result in smaller decay lengths.
But the lower momenta also
tend to yield larger opening angles of the daughter tracks
and therefore smaller uncertainties on $L$.
Thus smaller decay lengths are associated with larger weights.
\item In some vertex fitting programs,
the covariance matrices describing the errors on the reconstructed
track parameters are ``swum'' to the location of the fitted vertex
and a second fitting iteration is performed.
This appears to be a reasonable thing to do,
but such a fitting program assigns larger weights, on average,
to the $\tau$'s with larger decay lengths, leading to a bias
on the average decay length.
\end{itemize}}

\subsection{Detector alignment}
Tracking systems are calibrated and surveyed based on tracks
reconstructed in the data.
This procedure is not perfect;
after alignment, the average impact parameter $\langle d \rangle$
(which would ideally be zero)
can vary with $\theta$ and $\phi$ by $10\mic$ or more.
The $\tau$ lifetime measurement is, however,
based on impact parameters on the order of $c\life = 87\mic$.
How can the experiments claim systematic uncertainties related
to detector alignment as small as $0.1\%$?

A conjecture, put forth by ALEPH~\cite{alephalign},
provides some insight.
They theorize that the effects of $d$ offsets cancel, to first order,
if there are no azimuthal holes in the acceptance of the tracking
system.
To illustrate this point, I present the preliminary results
of a simple Monte Carlo study.
Simulated $\ee\To\tautau$ events at
$\sqrt{s} = 91.2 \gev$ were generated, and
a sample of $500\,000$ three-prong decays was selected with
reasonable cuts on the momentum and polar angle of the
daughter tracks.
Rather sterile conditions were maintained for the experiment:
Gaussian tracking errors were generated, with
$\sigma_d = 30\mic$,
$\sigma_{\phi} = 0.2\mrad$,
and $\langle \delta d \cdot \delta\phi \rangle =
0.9 \sigma_d \sigma_{\phi}$.
A two-dimensional vertex fit was performed for each decay, and
no errors on the $\tau$ production point or the $\tau$ direction
were introduced in the calculation of the decay length.
Impact parameter offsets were then applied as a function of $\phi$,
and the fits were repeated,
to study the effect on the decay length bias
$\langle L_{\rm rec} - L_{\rm true} \rangle$.
I tried four different $d$ offset configurations, as described below.
\begin{figure}[t]
\begin{center}\mbox{%
\rlap{\epsfxsize=67.5mm\epsffile{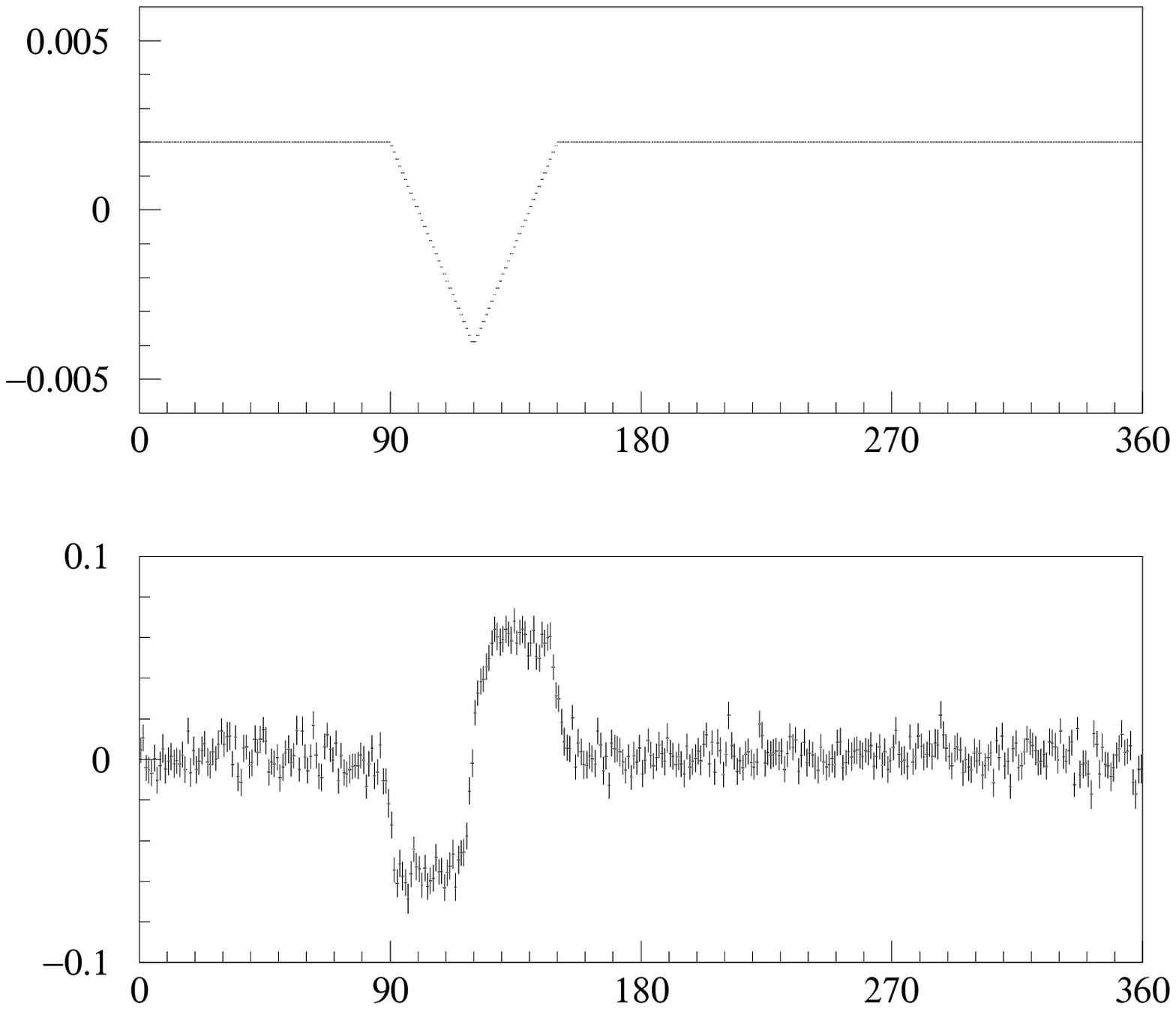}}%
\rlap{\epsfxsize=67.5mm\epsffile{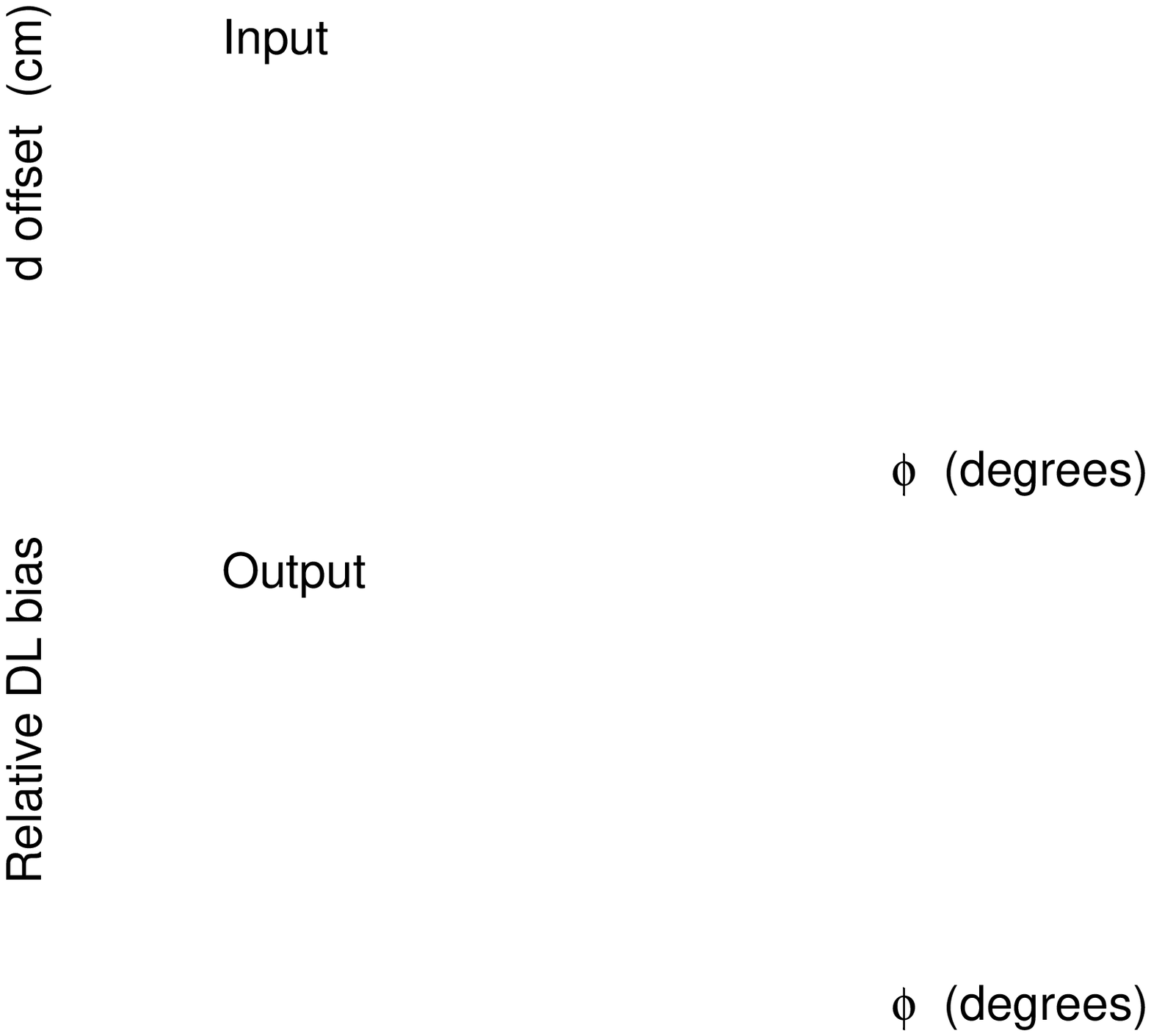}}%
\rule{75mm}{0mm}}\end{center}\par\vspace{-10mm}
\caption{Monte Carlo study of impact parameter offsets, Experiment~B
(uniform offset plus one excursion).
The ``Input'' plot shows the offsets 
applied to the impact parameters $d$
as a function of $\phi$.
The ``Output'' plot shows the resulting relative bias on the
reconstructed decay length,
$B = \langle L_{\rm rec} - L_{\rm true} \rangle /
\langle L_{\rm true} \rangle$
as a function of $\phi$.}
\label{f:offsetb}
\end{figure}
\begin{figure}[t]
\begin{center}\mbox{%
\rlap{\epsfxsize=67.5mm\epsffile{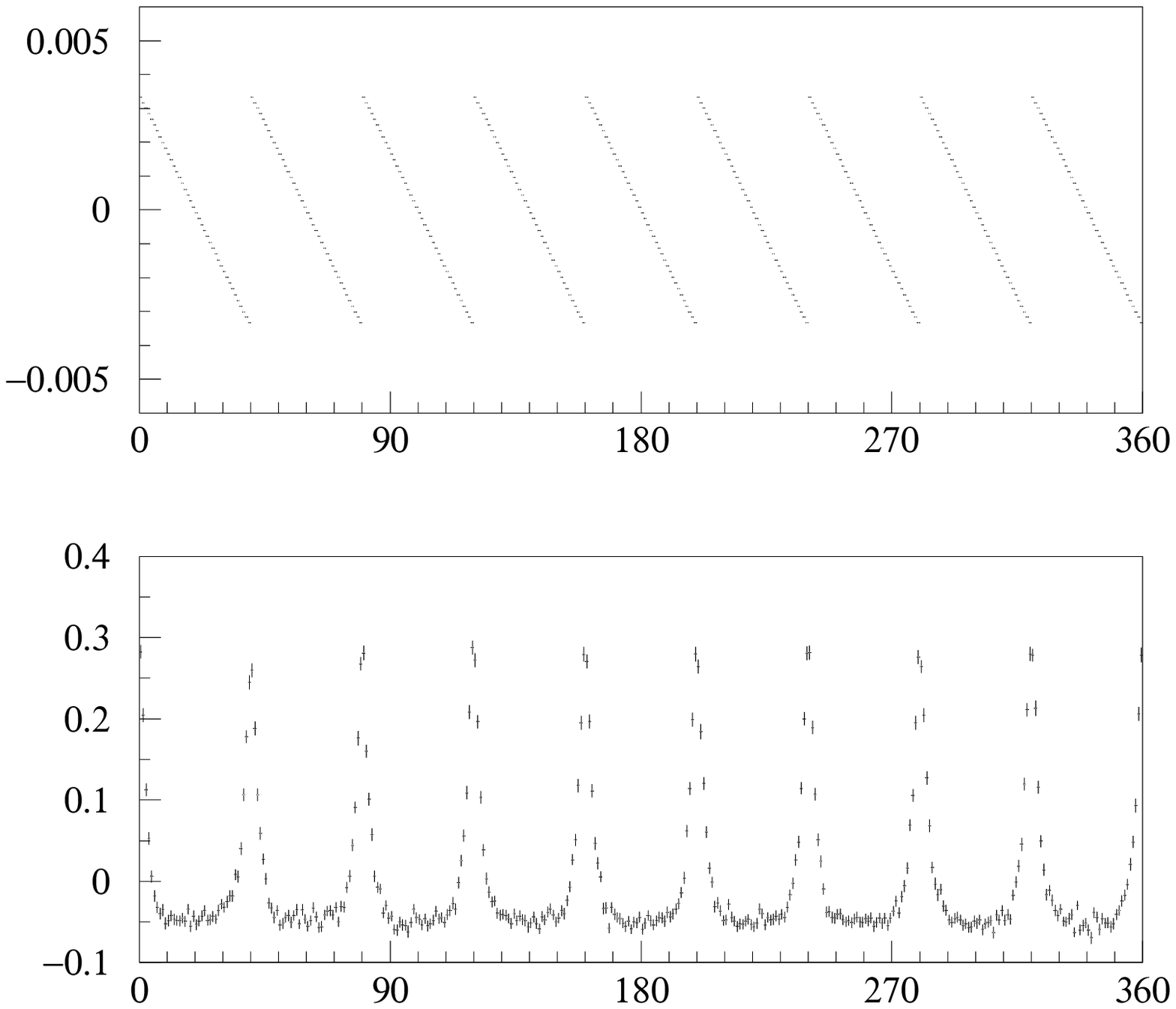}}%
\rlap{\epsfxsize=67.5mm\epsffile{teenal_labels.ps}}%
\rule{75mm}{0mm}}\end{center}\par\vspace{-10mm}
\caption{Monte Carlo study of impact parameter offsets, Experiment~C
(radial shift of silicon vertex detector wafers).}
\label{f:offsetc}
\end{figure}
{\mylists
\begin{itemize}
\item[A.] No $d$ offsets.
When no systematic offsets are applied to the impact parameters,
the average bias is
$B = \langle L_{\rm rec} - L_{\rm true} \rangle /
\langle L_{\rm true} \rangle
= (+0.16 \pm 0.04)\%$,
reflecting the small positive bias due to the correlation
of the $d$ and $\phi$ errors.
The rule of thumb is that the relative decay length bias
is roughly equal to the ratio
of the detector-induced correlation of $d$ and $\psi$ to the
lifetime-induced correlation.
In this case the detector-induced correlation is
$\langle \delta d \cdot \delta\phi \rangle = 0.0054\mic$,
while the lifetime-induced correlation is roughly
$\langle d \cdot \psi \rangle = (c\life)(1/\gamma) = 3.4\mic$,
so the ratio is $0.0054/3.4 = 0.0016$,
which is comparable (!) to the observed offset.

\begin{figure}[t]
\begin{center}\mbox{%
\rlap{\epsfxsize=67.5mm\epsffile{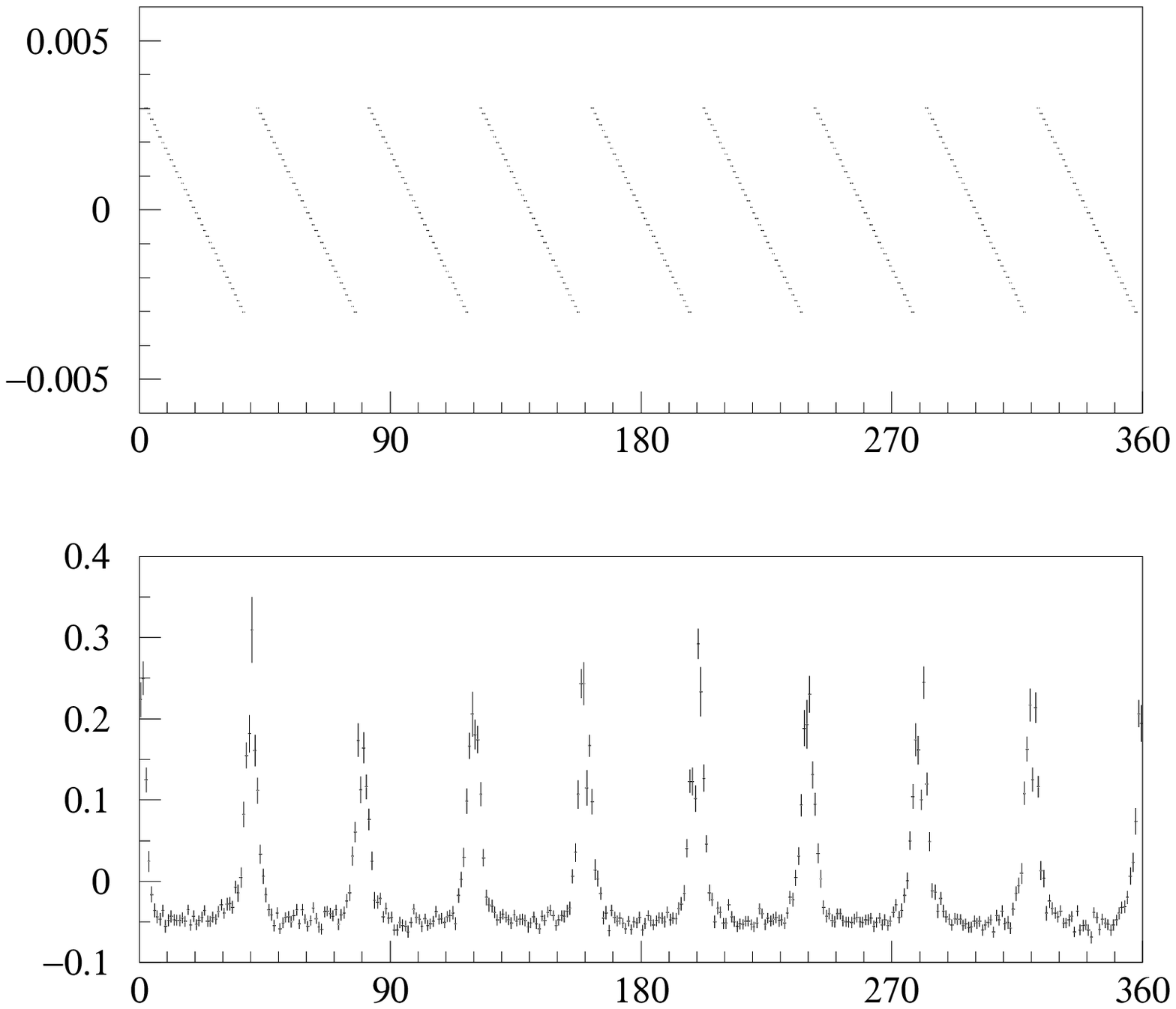}}%
\rlap{\epsfxsize=67.5mm\epsffile{teenal_labels.ps}}%
\rule{75mm}{0mm}}\end{center}\par\vspace{-10mm}
\caption{Monte Carlo study of impact parameter offsets, Experiment~D
(broken azimuthal acceptance).}
\label{f:offsetd}
\end{figure}
\item[B.] Uniform offset plus one excursion.
The ``Input'' plot in
Fig.~\ref{f:offsetb} shows the applied
$d$ offset of $+20\mic$ everywhere,
plus a triangular excursion of amplitude $-60\mic$ in one region.
The ``Output'' plot shows that the local bias $B$
is essentially the derivative of the
input function with respect to $\phi$,
with the sharp edges smoothed out over an angular scale
corresponding to the typical opening angle of the $\tau$ decays.
In particular, the slope of the offset function in the region
of the excursion is 
${\pm}(60 \mic)/(30^{\circ}) = {\pm}115\mic$ 
(converting the degrees to radians).
This quantity is equal to ${\pm}6.1\%$ of the mean decay length in the
$xy$ projection,
whereas the maximum observed bias in the output plot
is about ${\pm}6\%$.
In spite of the large local biases, the global average bias is
$B =  (+0.17 \pm 0.04)\%$,
i.e., unchanged from Experiment~A.
If the acceptance is unbroken in $\phi$, and the bias is
the derivative of the input function,
then the average bias is proportional to the integral of the derivative,
which is zero for {\it any\/} input offset function.

\item[C.] Radial shift of silicon vertex detector faces.
Here I consider a one-layer vertex detector
with nine flat faces at a radius of $6\cm$.
I suppose that the silicon wafers are shifted away from the origin
by $100\mic$ with respect to their assumed locations.
I then make the crude approximation that this shift has no effect
on the reconstructed track directions and simply introduces an
offset on $d$ given by $(-100\mic)\sin\alpha$,
where $\alpha$ is the azimuthal angle of incidence of the
track on the wafer.
This scenario corresponds to the input function shown in
Fig.~\ref{f:offsetc}.
Again the output plot looks like the derivative of the input:
the local bias has positive $\delta$ functions 
(again smeared due to the opening angles of the $\tau$ decays)
in the regions where the daughter tracks do not all pass through
the same vertex detector face
and a uniform negative value elsewhere.
The bias is locally as large as $+28\%$.
Nevertheless, the global average bias is unchanged:
$B =  (+0.14 \pm 0.04)\%$.

\item[D.] Broken azimuthal acceptance.
This experiment is the same as Experiment~C, 
except that $2^{\circ}$ azimuthal gaps are 
introduced between adjacent faces of the vertex detector.
I reject $\tau$ decays in which one or more of the daughter tracks
passes through a gap.
Naturally, this has no effect on the output plot (Fig.~\ref{f:offsetd})
except near the positive spikes;
some of the events that had a large positive bias in Experiment~C
are now removed from the sample.
The resulting negative bias on the global mean decay length is huge:
$B =  (-3.73 \pm 0.04)\%$.
If the gaps had been wider,
such that the three daughter tracks in selected events 
always pass through the same face of the vertex detector, 
the bias on the mean $xy$ decay length would have been simply
$-100\mic$ (the radial shift of the faces)
or $-5.3\%$.
\end{itemize}}

These experiments show that $d$ offsets 
due to alignment and calibration errors
in tracking systems have little effect on the measured lifetime
in the DL method,
as long as the azimuthal acceptance is unbroken.
In fact it is straightforward to prove that the same result
holds for the IPD method.
But these conclusions rely on the assumption that the
weighting of the events in the lifetime averaging procedure 
is independent of $\phi$.
In a realistic situation where the smearing related to the
luminous region is significant (not at SLD!)
and depends on $\phi$,
the (reweighted) integral of the derivative of the offset function
is, in general, not equal to zero.

It should be mentioned that correlated offsets in $d$ and $\phi$
can yield a bias even when the azimuthal acceptance is unbroken.
It is straightforward to measure the $d$ offsets versus $\theta$ and
$\phi$ from Bhabha, dimuon, or $\qqbar$ events.
Corrections may then be applied to the $\tau$ data.
On the other hand, offsets in $\phi$ are difficult to measure;
the systematic uncertainty on the lifetime should allow for
a range of possibilities.

Systematic offsets in impact parameter and direction may also be present
in the $rz$ view.
These offsets, affecting methods such as DL and 3DIP, which do not operate
solely in the $xy$ projection,
are difficult to measure.
Moreover, there is no bias cancellation rule for such offsets because
the unbroken acceptance and periodic boundary conditions do not
apply in $\theta$ as they do in $\phi$.

Finally, it is interesting to note that the absolute scale of the
lifetime measurements is set by the impact parameter scale,
which depends on the detector dimensions.
In experiments with microstrip or pixel vertex detectors,
it is the pitch of those detector elements that counts,
not the radii of the vertex detector layers.

\section{Treatment of biases}
With some methods, the $\tau$ lifetime measurement must be
``calibrated'' by means of Monte Carlo events.
For example, the interpretation of 
the impact parameter sum distributions studied in the MD/MIPS methods 
is based on simulated events.
In other methods (e.g., DL and IPD) 
there is a simple geometric relation
between the observables and the mean lifetime,
so, to first order, no calibration is needed.
In such cases the Monte Carlo is normally used 
to check for ``possible'' biases in a measurement;
a sample of events with known generated lifetime
is passed through the analysis, 
and the reconstructed mean lifetime is compared to the input value.
The experimenters may then choose one of two valid approaches:
{\mylists
\begin{enumerate}
\item If $\tau_{\rm output}$ is significantly different from
$\tau_{\rm input}$, 
apply the difference as a correction to the lifetime measured
in the data.
\item Always apply the difference
as a correction to the lifetime measured in the data.
\end{enumerate}}
Approach~1 is used by most experiments.
It turns out that between 1985 and 1993 
the difference $\tau_{\rm output} - \tau_{\rm input}$
was not subtracted
in eight published measurements of $\life$ with the DL method.
In all eight cases, $\tau_{\rm output}$
was greater than $\tau_{\rm input}$~\cite{srw}.
This observation is experimental evidence that 
common biases are present in all DL measurements.

Corrections must be applied for all biases before
a valid world average can be calculated.
I would encourage experimenters to apply those corrections themselves,
but to go far beyond Approach~2.
We can and should identify and measure 
each {\it individual\/} source of bias
in our analyses by means of Monte Carlo events.
Here is an example from the IPD method.
The bias that results from radiative events in which
the $\tau^{+}$ and $\tau^{-}$ are {\it not\/} back to back in the
$xy$ projection can be evaluated by calculating $\Delta\phi$
(see Section~\ref{ss:ipd})
from Monte Carlo truth information,
with and without a correction for the acoplanarity of the $\tau$'s,
and comparing the fitted lifetimes in the two cases.
Such a technique yields a measurement of this one bias
with very good statistical precision,
and the reliability of the simulation can be checked by
searching for $\ee\To\gamma\tautau$ events in data and Monte Carlo.

A series of measurements of this type can be devised,
using various pieces of information from the Monte Carlo truth,
in order to evaluate each contribution to
$\tau_{\rm output} - \tau_{\rm input}$.
We can rigorously evaluate the systematic uncertainty              
on the lifetime measurement
only by understanding the magnitude of each bias contribution.

\section{History of the $\tau$ lifetime}
\begin{figure}[t]
\begin{center}
\mbox{\epsfxsize=55mm\epsffile{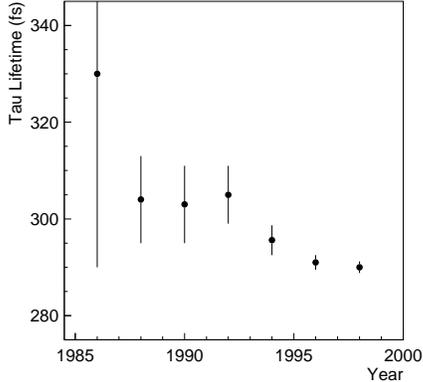}}%
\end{center}\par\vspace{-15mm}
\caption{RPP world average $\tau$ lifetime value
versus year~\cite{pdg}.}
\label{f:history}
\end{figure}

The world average $\tau$ lifetime values evaluated by 
the Particle Data Group since 1986
are plotted in Fig.~\ref{f:history}.
A fairly steady decline is observed in these averages.
In my opinion, three factors contribute to this decline:
statistics,
unsubtracted biases,
and ``other effects.''
As evidence for the presence of ``other effects,''
the $\tau$ lifetime average
in the 1992 Review of Particle Properties~\cite{pdg}
had $\chi^2 = 2.0$ for 10 degrees of freedom,
corresponding to a confidence level of $0.9962$.

\section{Thoughts on the next method}
Here are a few ideas about a possible ultimate method
for analyzing \oneone\ topology events.
We may be able to squeeze a little more out of the \oneone\ 
(and perhaps \onethree) topology events 
if we created a method that takes into account all of the
available information:
energies and directions of charged and neutral particles,
charged particle identification,
impact parameters in $r\phi$ and $rz$ views, and
the position and size of the luminous region.

The method should
transform this information into two or three variables
from which the lifetime is extracted.
It should take into account the known $\tau$ decay dynamics
and allow for initial and final state radiation.
Furthermore, I believe we will not be able to make
substantial advances in precision unless 
the new method allows the impact parameter resolution 
to be fitted from the $\tau$ data itself, 
as in the 3DIP method.

\section{Outlook and conclusions}
In conclusion, the world average $\tau$ lepton lifetime
is
\[ \life = 290.5 \pm 1.0 \fs, \]
and the $\chi^2$ of the average looks healthier.
The LEP experiments dominate the world average.

It will be a challenge to achieve the next factor of 8
improvement in precision on $\life$.
We will need to rely on the large statistics of CLEO and the b factories,
with considerable care to understand and reduce systematic errors
related to the fitting procedures, tracking resolution,
and backgrounds.

\section*{Acknowledgements}
I would like thank the organizers,
particularly Toni Pich and Alberto Ruiz,
for giving me the chance to participate in the workshop.
It was a pleasure to visit Santander and the surrounding area,
and the arrangements, accomodations, and food were excellent!
I would also like to thank 
Attilio Andreazza,
Roy Briere,
Auke-Pieter Colijn,
Mourad Daoudi,
Jean Duboscq,
Wolfgang Lohmann,
Duncan Reid,
Patrick Saull,
Abe Seiden, and
Randy Sobie
for providing information for this talk.

\end{document}